\documentclass[preprint,aps,pre,amsmath,eqsecnum,showpacs,floatfix]{revtex4}
\usepackage{epsfig}
\usepackage{amssymb,amsmath}

\def\be{\begin{equation}}
\def\ee{\end{equation}}

\def\LDC{\langle\!\langle}
\def\RDC{\rangle\!\rangle}
\def\CH{\mathcal{H}}

\def\bea{\begin{eqnarray}}
\def\eea{\end{eqnarray}}

\def\ba{\begin{array}{l l}}
\def\ea{\end{array}}

\usepackage{epsfig}
\usepackage{amssymb,amsmath}
\usepackage{graphicx}
\begin{document}

\title{EFFECT OF BIQUADRATIC EXCHANGE  ON THE PHASE DIAGRAM OF   A SPIN-1 TRANSVERSE XY MODEL WITH SINGLE-ION ANISOTROPY}

\author{ I. Rabuffo$^{1}\footnote{Corresponding author \\ E-mail address: rabuffo@physics.unisa.it}$, L. De~Cesare$^{1}$, A. Caramico~D'Auria$^{2}$, M.~T. Mercaldo$^{1}$}

\affiliation{$^1$ Dipartimento di Fisica ``E. R. Caianiello'',
Universit\`a di Salerno and CNISM, Unit\`a di Salerno,  I-84084
Fisciano (Salerno), Italy} \affiliation{$^2$ Dipartimento di
Fisica, Universit\`a di Napoli Federico II, I-80125
Napoli, Italy  }

\begin{abstract}
The two-time Green function method is employed to explore the effect of the biquadratic exchange interaction on the phase diagram of a $d$-dimensional spin-1 transverse XY model with single-ion anisotropy  close to the magnetic-field-induced quantum critical point. We work at level of the Anderson-Callen-decoupling-like  framework for both easy-plane and easy-axis single-ion anisotropy. The structure of the phase diagram is analyzed with analytical estimates and numerical calculations by adopting  Tyablikov-like decouplings for the exchange higher order Green functions in the equation of motion. For dominant bilinear short-range exchange interaction  the structure of the phase diagram close to the quantum critical point remains qualitatively the same of that in absence of biquadratic interaction including reentrant critical lines. When the biquadratic exchange becomes increasingly dominant its role appears more effective and tends to reduce or destroy the reentrant character of the critical lines.      
\end{abstract}

\pacs{64.70.Tg  75.10.Jm  75.40.-s 05.70.Fh}

\maketitle

\section{Introduction}
\label{sec1}

The combined role played by biquadratic exchange (BQE) and single-ion anisotropy (SIA) in magnetic models has attracted great interest for several decades  with a renewed attention in the recent years. Several analytical and numerical studies have been achieved for spin-$S \geq 1$ magnetic systems with Heisenberg and XXZ symmetry by using different approximation schemes (see, for instance, \cite{Chen, Micn, Chad}),  including the powerfull two-time Green Function (GF) method \cite{tya},\cite{nolting}. This method has had a great impact in condensed matter physics and specially in quantum theory of magnetism from its diffusion in the scientific literature after the famous review article by Zubarev \citep{zubarev} almost sixty years ago. In particular, it has received recently a great attention for understanding the thermodynamic properties of advanced magnetic materials with general non-collinear magnetic structures as the helimagnetic thin films \citep{diep1, diep2, diep3}. From these investigations, a very rich structure of the phase diagram emerges providing a remarkable tool for understanding the complex magnetic properties of several innovative materials \cite{Ega, Hirs, Wis, Wis2, And, Bra}, also involving quantum phase transitions (QPTs) \cite{sachdev,noi07,noi10}.
The effect of the BQE interactions on the finite temperature phase transitions in systems with XY symmetry, also in presence of a SIA, has received less attention. 
A quantum spin-1 anisotropic XY model (really an XZ model in a non-conventional geometry) with BQE and an in-plane magnetic field has been investigated  in Ref.\cite{Chad2}. Here, the two-time GF technique, limited to the random phase approximation (RPA) \cite{tya},\cite{nolting} for higher order exchange GFs and with undecoupled  GFs for single site spin operators,  is employed to obtain the critical temperature as a function of the BQE and the SIA parameters.  Besides, Monte Carlo simulations have provided a lot of results for the classical XY model by variation of the bilinear to biquadratic exchange interaction ratio \cite{Chen2,Chen3,Zuk, Zuk2}.
However, to our knowledge, studies about the role played by competition of BQE and SIA on the phase diagram of magnetic systems which exhibit a quantum critical point (QCP) are lacking at the present time. 
In this context, the simple XY model in a transverse magnetic field \cite{Dut} is of particular interest as a paradigmatic spin model capturing the essential physics of the so-called planar magnets  and may be a good laboratory of investigation. It has received much attention for several decades from both theoretical and experimental points of view and, at the present time, its basic static and dynamic quantum critical properties are well established \cite{sachdev}, \cite{Dut}. Recent studies  have focused on quantum criticality of the model with SIA but in absence of BQE and several interesting findings have been established by employing different methods depending on the range of values of the anisotropy parameter \cite{lima}-\cite{noi16}. 

The present paper is devoted to explore, by using  the two-time GF method, the effect of the BQE on the phase diagram of a $d$-dimensional spin-1 transverse XY  (TXY) ferromagnetic model with both easy-axis and easy-plane SIA,  working at level of the Anderson-Callen decoupling (ACD)\cite{ACD}   scheme. We adopt  Tyablikov-like decouplings (or RPA) for the exchange higher order Green functions in the equation of motion (EM).  For short-range exchange interactions, the structure of the phase diagram  is then analyzed for dimensionalities $d > 2$ by means of analytical estimates and numerical calculations.

  The paper is organized as follows. In Sec. II we introduce  the spin model of interest and present the EM for the appropriate two-time GF. Sec. III is devoted to the ACD treatment focusing on the role played by the BQE  on the phase diagram around the magnetic-field-induced  QCP by variation of the exchange and the SIA parameters. Analytical estimates for the critical lines  ending in the QCP are obtained for short-range interactions.  Numerical findings are presented in Sec.IV. Finally, in Sec.V conluding remarks are drawn.    

\section{Spin model and  two-time Green function framework }
\label{s2}

The spin model we consider here  is described by the Hamiltonian
\bea
\label{HAM}
\CH &=& -\frac12 \sum_{i,j=1}^N \left\{J_{ij} (S_i^x S_j^x +S_i^y S_j^y)
+I_{ij} (S_i^x S_j^x + S_i^y S_j^y)^2 \right\} \nonumber \\
&-& D \sum_{i=1}^N(S_i^z)^2 -H  \sum_{i=1}^N S_i^z.%
\eea
where $S_i^\alpha (\alpha=x,y,z)$ are the  components  of the
 spin-1 vector $\vec{S}_i$ located at site $i$ of a $d$-dimensional hypercubic lattice with $N$ sites and unit lattice spacing.
Here $J_{ij}$ and $I_{ij}$  (with $J_{ii}=I_{ii}=0$) are the ferromagnetic  bilinear and biquadratic  exchange couplings,  $D$ denotes the easy-axis $(D>0)$ or easy-plane  $(D<0)$ SIA parameter and $H$ is the applied magnetic field along the $z$-direction in the spin-space. Through this paper we focus on the short-range interactions but the case of long-range ones can be similarly analyzed. 

We  introduce the retarded two-time GF
\bea
\label{GF1}
G_{ij}(t-t')&=& - i \theta(t-t') \langle [S_i^+(t-t'),
S_j^-]\rangle  \nonumber \\
&\equiv&\LDC S_i^+; S_j^- \RDC  ,
\eea
where $\theta(x)$ is the step function, $[  ,  ]$ denotes a commutator, $A(t)$ is the Heisenberg representation of the operartor  $A$, $\langle \cdots \rangle$   stands for an equilibrium average at temperature $T$  and $S_i^{\pm}=S_i^x\pm i S_i^y$.
The   EM for the time Fourier transform
$G_{ij}(\omega)=\int_{-\infty}^\infty dt\, e^{-i\omega t}G_{ij}(t) 
\equiv \LDC S_i^+|S_j^-\RDC_\omega$, after 
some spin-commutator algebra, reads

\bea
\label{eqmot}
&&(\omega-H)G_{ij}(\omega) =
2m \delta_{ij}-\sum_h \left\{ J_{ih} \LDC S_i^z S_h^+;S_j^-\RDC_\omega
+\frac{1}{2} I_{ih}\left[ \LDC (S_h^+)^2 A_i^-;S_j^-\RDC_\omega +2 \LDC A_i^+;S_j^-\RDC_{\omega}  \nonumber \right.\right. \\  && \left.\left. - \LDC (S_h^z)^2 A_i^+; S_j^-\RDC_\omega + \LDC S_i^z S_i^+;S_j^-\RDC_\omega \right]\right \} + D\LDC A_i^+;S_j^-\RDC_\omega .
\eea
Here
$m =  <S_i^z>$   is  the longitudinal  magnetization per spin and
\be
\label{A}
A_i^\pm = S_i^z S_i^\pm + S_i^\pm S_i^z.
\ee
As we see the EM for $G_{ij}(\omega)$  is not in a closed form since higher-order GFs occur. Then one is forced to adopt proper decouplings in order to close the infinite chain of EMs which determine $G_{ij}(\omega)$. Here we use Tyablikov-like decouplings \cite{tya},\cite{nolting}, for the three- and five-spin exchange GFs, which neglect completely the correlations between the longitudinal and trasversal spin components at different lattice sites. For the remaining SIA-like terms  we employ the usual ACD \cite{ACD}.

\section{The Anderson- Callen decoupling framework}
\label{sec3}

As mentioned in the previous section, we begin to  decouple the higher order GFs in Eq. \eqref{eqmot}  as follows.
For exchange terms  we assume the Tyablikov-like decouplings,  

\bea
\label{decoupling}
&&\LDC S_i^z S_h^+|S_j^-\RDC_\omega \simeq \langle S_i^z\rangle\LDC S_h^+|S_j^-\RDC_\omega = m G_{hj} (\omega),\nonumber \\
&&\LDC (S_h^+)^2 A_i^-;S_j^-\RDC_\omega \simeq \langle (S_h^+)^2\rangle\LDC A_i^-;S_j^-\RDC_\omega = 0, \nonumber \\
&&\LDC (S_h^z)^2 A_i^+; S_j^-\RDC_\omega \simeq\langle(S_h^z)^2\rangle\LDC A_i^+;S_j^-\RDC_\omega.
\eea
The remaining SIA-like GF $\LDC A_i^+; S_j^-\RDC_\omega$ will be treated by using the ACD \cite{ACD}
\be
\label{SIALIKE}
\LDC A_i^+; S_j^-\RDC_\omega\simeq 2\langle S^z_i\rangle\left(1-\frac{\langle S_i^-S_i^+ + S_i^+S_i^-\rangle}{4•}\right)G_{ij}(\omega)=m\langle (S^z)^2\rangle G_{ij}(\omega),
\ee
where use has been made of the kinematic identity $S_i^- S_i^+ + S_i^+S_i^-  = 2(2 - (S_i^z)^2)$.

Working in the wave-vector-frequency $(\vec{k},\omega)$-space with $G_{ij}(\omega)=\frac 1 N \sum_{k} e^{i\vec{k}\cdot(\vec{r}_i-\vec{r_j})}G(\vec{k},\omega)$ and $X(\vec{k})=\sum_{\vec{r}} X(|\vec{r}|)e^{i\vec{k}\cdot\vec{r}}\quad (X=J,I)$,  $\vec{k}$ ranging in the first Brillouin zone (1BZ),  EM \eqref{eqmot} reduces to an algebraic equation for $G(\vec{k},\omega)$  which has the polar solution

\be
\label{Gpolar}
G(\vec{k},\omega)= \frac {2m}{\omega-\omega(\vec{k})}.
\ee
Here, 
\be
\label{OMEGA}
\omega(\vec{k})= \omega_0+ m \left(J(0)-J(\vec{k})\right),
\ee
defines the "energy spectrum" of the undamped spin excitations with energy gap 

\be
\label{OMEGA0}
\omega_0\equiv \omega(0)= H - m\left\{\left(J(0)+\frac{I(0)}{2}\right)-\langle(S^z)^2\rangle\left[D-\frac{I(0)}{2}(2-\langle(S^z)^2\rangle)\right]\right\}.
\ee
Bearing in mind the kinematic identity 
\begin{equation}
\label{kinematic}
(S^z_i)^2= 2 - S_i^z-  S_i^-S_i^+,
\end{equation}
and the relation

\be
\label{spectral}
\langle S^-_i S^+_i\rangle= 2m \Phi,
\ee
arising from "spectral theorem" \cite{noi06},\cite{noi08} for $G_{ij}(\omega)$, we get 
 
\be
\label{essezeta}
\langle (S^z_i)^2\rangle = 2- m(1+2\Phi),
\ee
where

\be
\label{phi}
\Phi=\frac{1}{N} \sum_{\vec{k}}\frac {1}{e^{\omega(\vec{k})/T}-1}\stackrel{ N\to\infty }{\rightarrow}\int_{(1BZ)}\frac{d^dk}{(2\pi)^d}\frac{1}{e^{\omega(\vec{k})/T}-1}.
\ee
Hence Eq. \eqref{OMEGA0} becomes

\be
\label{OMEGA0bis}
\omega_0 =  H - m\left\{J(0) - D\left[2-m(1+2\Phi)\right]+\frac{I(0)}{2}\left[2-(1-m (1+2\Phi))^2\right]\right\}.
\ee

In the present scheme, we need  to determine an expression for $m$ in terms of the correlation functions related to the GF. This can be achieved by using the Callen method \cite{CallenM} based on the introduction of the parametric GF   $ \LDC S_i^+; e^{a S_i^z} S_j^-\RDC$ where $a$ is an auxiliary parameter  to be set to zero at the end of calculations.
For our spin-1 model the proper relation for $m$ is  \cite{noi14a}, \cite{noi14b},\cite{noi13}.

\be
\label{emme}
m= 1- \Phi + \frac{3}{(1+\frac{1}{\Phi})^3 - 1}
\ee
with the asymptotic expansions
\bea
\label{BDIPHI}
m\simeq
\left\{
\ba
1- \Phi +3\Phi^3+\cdots\quad \quad  \quad \quad, \quad   \quad  \Phi\ll 1
\\
\\
\frac 2 {3\Phi}\left(1-\frac{1}{2\Phi}+\frac{1}{6\Phi^2•}+\cdots\right)\quad \quad,\quad \quad \Phi\gg 1.
\ea
\right.
\eea
Notice that, $m \rightarrow 0$ when  $\Phi\rightarrow \infty$  but  $m\Phi\rightarrow\frac{2}{3}$.

Eqs.\eqref{phi}-\eqref{emme} constitute a set of self-consistent equations for exploring the thermodynamic properties and the phase diagram of the model \eqref{HAM} with competing exchange couplings and single-ion anisotropy.

The quantity of  interest for our purposes is the transverse susceptibility defined by
\be
\label{transvsusc}
 \chi_{\perp}=-G(0,0)=\frac{2m}{\omega_0},
 \ee 
 the thermodynamic stability $(\chi_{\perp}>0)$ requiring  that $\omega_0\geq 0$. Then, the possible critical points are determined by assuming   $\omega_0 = 0^+$ and hence by solving the equation
 
\be
\label{critical}
H - m\left\{\left (J(0)+\frac{I(0)}{2}\right) - \left[2-m(1+2\Phi_c)\right]\left[D-\frac{I(0)}{2}m(1+2\Phi_c)\right]\right\}=0,
 \ee 
with  (see Eqs \eqref{OMEGA0} and \eqref{phi})

\be
\label{phic}
\Phi_c= \Phi\mid_{\omega_0=0}\quad
\stackrel{ N\to\infty }{=}\int_{(1BZ)}\frac{d^dk}{(2\pi)^d}\frac{1}{\exp[\frac{1}{T}m(J(0)-J(\vec{k}))]-1}.
\ee
It is immediate to see that, at $T=0$, one has $\Phi_c=0$ implying the solution $m=1$ (full polarized ground state) and hence the existence of a magnetic-field-induced QCP with coordinates  $(H_c=J(0)+I(0)-D,\quad T=0)$ for any dimensionality. Of course, this happens only if $D< J(0)+I(0)$.  Notice that for $D=J(0)+I(0)$ the quantum critical field $H_c$ vanishes and the QCP is suppressed.
Previous result for $H_c$ suggests that the inclusion of the BQE  coupling in  the Hamiltonian \eqref{HAM} produces an increase of $H_c$ (for the conventional TXY model with SIA $H_c=J(0)-D$). 
For case of  short-range interactions and $d$-dimensional hypercubic lattices, the exchange interactions behave as $X(\vec{k})\simeq X(0)-Xk^2  \quad (X=J,I)$ as $k\rightarrow 0$ where $X(0)=zX$ and $z$ is the coordination number with $z=6,8,12$ for three-dimensional $sc$, $bbc$ and $fcc$ lattice, respectively.  So, a finite-temperature critical line  takes place only for dimensionalities that assure the convergency of the integral \eqref{phic}, i.e. olnly for  $d>2$, consistently with the Mermin-Wagner theorem \cite{MW} . To explore the structure of the phase diagram close to the QCP when the BQ coupling is active, we need numerical calculations. However, preliminary information can be extracted from Eqs. \eqref{phi}-\eqref{BDIPHI} in the regimes $\Phi_c\ll 1$ and $\Phi_c\gg 1$ where the critical line equation \eqref{critical} is found to have the asymptotic expressions

\bea
\label{ASYMPCRITICAL}
H\simeq
\left\{
\ba
H_c-[J(0)-I(0)-2D]\Phi_c - (\frac{I(0)}{2}+3D)\Phi_c^2 + O(\Phi_c^3),\quad \quad \Phi\ll 1
\\
\\
\frac {2} {3\Phi_c} \left [(J(0)+\frac{17}{18}I(0))-\frac{2}{3}D\right ]\left(1-\frac{1}{2\Phi_c}\right) + O(\frac{1}{\Phi_c^3}).\quad\quad\quad\quad\quad \Phi\gg 1.
\ea
\right.
\eea

Analytical estimates can be obtained providing explicit expressions of $\Phi_c$ as a  function of temperature in the previous asymptotical regimes. This can be achieved for $d>2$ assuming the low-wave-vector approximation $J(0)-J(\vec{k})\simeq Jk^2$ in the integral \eqref{phic}. Taking into account the expression \eqref{BDIPHI} for $m$ we obtain from Eq. \eqref{phic} the low-temperature behavior
\be
\label{FIcritico}
\Phi_c(T)\approx \frac {K_d}{2}\Gamma (d/2)\zeta (d/2)\left( \frac{T}{J}\right)^{d/2} + \frac{d}{2}\left[\frac {K_d}{2}\Gamma (d/2)\zeta (d/2)\right]^2\left(\frac{T}{J}\right)^d+O\left(\left(\frac{T}{J}\right)^{3d/2}\right),
\ee
where $K_d=2^{1-d}\pi^{-d/2}/\Gamma (d/2)$ and $\Gamma(x)$ and $\zeta (x)$ are the Euler gamma and the Riemann zeta functions. Then, Eq. \eqref{ASYMPCRITICAL} provides the required critical line equation in the ($H,T$)-plane critical line close to the QCP

\bea
\label{CRITICALHT}
&& H\simeq H_c -\frac {K_d}{2}\Gamma (d/2)\zeta (d/2)\left[J(0)+I(0)-2D\right]\left( \frac{T}{J}\right)^{d/2}
- \frac{1}{2}\left[\frac {K_d}{2}\Gamma (d/2)\zeta (d/2)\right]^2 \times
\nonumber  \\  
&& \times \left\{d\left[J(0)+I(0)-2D\right]+ (I(0) + 6D)\right\}\left(\frac{T}{J}\right)^d + O\left(\left (\frac{T}{J}\right)^{3d/2}\right).
\eea
Solving this  equation with respect to $H$ or $T$ one obtains the representations $H_c(T)$ or $T_c(H)$.
By inspection of Eq. \eqref{CRITICALHT} the following scenario for $d>2$ emerges.
In the low-temperature regime $T/J \ll 1$ a conventional phase diagram in the $(H,T)$-plane takes place for $D< (J(0) + I(0))/2$,  quite similar to that of TXY model in absence of BQ interaction \cite{noi14a, noi14b, noi15, noi16}, \cite{noi13},\cite{noi17}, \cite{mtm17}. For $(J(0) + I(0))/2 < D < J(0) + I(0)$, one has $J(0) + I(0) - 2D < 0$ in Eq. \eqref{CRITICALHT} suggesting the existence of critical lines with reentrant behavior close to the QCP. This means that, in our approximation, reentrant phenomena in the phase diagrams occur only within a limited range of the easy-axis SIA parameter. Besides, in the easy-axis window $0<D<(J(0) + I(0))/2$ and in the easy-plane regime $D<0$ no reentrant critical lines exist. As shown in Sec. IV, numerical calculations confirm this   estimated scenario. In both cases, the shift exponent (defined  by $H_c(T) \approx H_c(0) - aT^{\psi}$ as $T\rightarrow 0)$ is $ \psi = d/2 $.

When $D = (J(0) + I(0))/2$ the term in $T^{d/2}$ vanishes and the critical line equation reads
\be
\label{CROSSCRITICAL}
H_c(T)\simeq H_c -\frac{1}{2}\left[\frac {K_d}{2}\Gamma (d/2)\zeta (d/2)\right]^2\left[3J(0)+4I(0)\right]\left( \frac{T}{J}\right)^{d} + O\left(\left(\frac{T}{J}\right)^{3d/2}\right).
\ee
This equation describes a regular (non-reentrant) critical line, with shift exponent $\overline{\psi}=2\psi=d$, signaling the crossover between regular critical lines and reentrant ones by increasing the SIA parameter $D$.
Eqs. \eqref{BDIPHI} and \eqref{FIcritico} predict that, in any case, the magnetization per spin along the critical line increases towards the full saturation value $m=1$ as $T\rightarrow 0$ with behavior

\be
\label{emmeci}
m_c(T)\simeq 1-\frac {K_d}{2}\Gamma (d/2)\zeta (d/2)\left(\frac{T}{J}\right)^{d/2}-\frac{d}{2}\left[\frac {K_d}{2}\Gamma (d/2)\zeta (d/2)\right]^2\left(\frac{T}{J}\right)^{d} + O\left(\left(\frac{T}{J}\right)^{3d/2}\right).
\ee
Notice that, at our level of approximation, along the critical lines the magnetization per spin as function of $T/J$ does not depend on $I$ and $D$.

A convenient procedure to determine $T_c(H)$ is to write Eq. \eqref{CRITICALHT} in the form 

\be
\label{CONVENIENTFORM}
A_d\left(\frac{T}{J}\right)^d + B_d\left(\frac{T}{J}\right)^{d/2} + \frac{H-H_c}{H_c}=0,
\ee
where

\be
\label{COSTA} 
A_d=\frac 12\left[\frac {K_d}{2}\Gamma (d/2)\zeta (d/2)\right]^2\frac{d[(J(0)+I(0)-2D]+(I(0) + 6D)}{H_c},
\ee

\be
\label{COSTB} 
B_d=\frac {K_d}{2}\Gamma (d/2)\zeta (d/2)\frac{(J(0)+I(0)-2D)}{H_c}.
\ee
This is  a quadratic algebraic equation in the variable $(T/J)^d$ providing, for $d>2$,

i) if $D < \frac{J(0)+I(0)} {2} \quad (B_d>0)$ one has the single physical solution 

\be
\label{SOLUTION1}
\frac{T_c(H)}{J}=\left(\frac{B_d}{2A_d}\right)^{2/d}\left\{\sqrt{1+\frac{4A_d}{B_d^2}\left(\frac{H_c-H}{H_c}\right)}-1\right\}^{2/d},
\ee
which, for $H$ sufficiently close to $H_c$, reduces to

\be
\label{SOLUTION2}
\frac{T_c(H)}{J}\simeq B_d^{-2/d}\left(\frac{H_c-H}{H_c}\right)^{2/d}\left\{1-\frac{2}{d}\frac{A_d}{B_d^2}\left(\frac{H_c-H}{H_c}\right)+O\left((\frac{H_c-H}{H_c})\right)^2\right\};
\ee

ii) if $ \frac{1}{2}[J(0)+I(0)] <D < J(0)+I(0)  \quad (B_d<0) $  the two physical solutions take place

\be
\label{SOLUTION3}
\frac{T^\pm_c(H)}{J}=\left(\frac{\mid B_d\mid}{2A_d}\right)^{2/d}\left\{1 \pm \sqrt{1- \frac{4A_d}{B_d^2}\left(\frac{H-H_c}{H_c}\right)}  \right\}^{2/d},
\ee
for $H_c\leq H \leq H^*$, where $H^*=\left(1+\frac{B_d^2}{4A_d}\right)H_c$  denotes the value of $H$ at which $T^+_c(H^*)\equiv  T^-_c(H^*) =  T^*=\left(\frac{\mid B_d\mid}{2A_d}\right)^{2/d}J$,  defining the double critical point $(H^*, T^*)$.
For $H$ sufficiently close to $H_c$ the solutions \eqref{SOLUTION3} reduce to

\begin{equation}
\label{ASYMPTC1}
\frac{T^+_c(H)}{J}=\left(\frac{\mid B_d\mid}{A_d}\right)^{2/d}\left\{1- \frac 2d \frac{A_d}{B_d^2}\left(\frac{H-H_c}{H_c}\right) + O\left((\frac{H_c-H}{H_c})\right)^2\right\},
\end{equation}
and
\begin{equation}
\label{ASYMPTC2}
\frac{T^-_c(H)}{J}=\mid B_d\mid^{-2/d}\left(\frac{H-H_c}{H_c}\right)^{2/d} \left\{1  + \frac 2d \frac{A_d}{B_d^2}\left(\frac{H-H_c}{H_c}\right) +  O\left((\frac{H_c-H}{H_c})\right)^2\right\}.
\end{equation}
Eqs. \eqref{ASYMPTC1} and \eqref{ASYMPTC2} show that also at $H=H_c$ we have two physical solutions: $T_c^-(H_c) = 0$ defining the QCP, and $T_c^+(H_c) = (\mid B_d\mid/A_d)^{2/d}J$ which locates an additional  point on the critical line. So, previous results suggest that at each $H$ in the interval $H_c \leq H \leq H^*$ two critical points occur,  with temperatures $T_c^-(H)$ which increases from zero to $T^*$ for $H$ increasing from $H_c$ to $H^*$, and $T_c^+(H)$ which decreases from $T_c^+(H_c)$  to $T^*$ with $H$ increasing from $H_c$ to $H^*$. For $H > H^*$ no critical point exists. Besides, for  each $H < H_c$ a single critical point occurs with  temperature $T_c(H) \equiv T_c^+(H)$ given by 
 \begin{equation}
 \label{TICIDIACCA}
 \frac{T_c(H)}{J} = (\mid B_d\mid/2A_d)^{2/d}\left[1 + (1 + \frac{4A_d}{B_d^2}(H_c - H)^{1/2})^{2/d}\right]
 \end{equation}
which decreases to $T_c^+(H_c)$  as $H \rightarrow H_c^- $;

iii) for $D = (J(0) + I(0))/2 \quad (B_d = 0)$ a critical point is selected with temperature 
\begin{equation}
\label{FRONTIER}
\frac{T_c(H)}{J}= A_d^{-1/d} \left[\frac{(H_c - H)}{H_c}\right]^{1/d}.      
\end{equation}

All these features are a clear evidence of  the existence of  reentrant critical lines for $ \frac{z}{2}(1+I/J) <D/J < z(1+I/J)$. In this range it may be useful to define the dimensionless quantity  $\delta = (H^* - H_c)/J$ which  measures the maximum amplitude of reentrances, for each fixed  $I/J$ and $D/J$. Estimates for $\delta$ follow immediately from the definitions of $H^*$ and $H_c$ using Eqs. \eqref{COSTA}, \eqref{COSTB}.  For cubic lattices  one finds 
\begin{equation}
\label{delta}
\delta=\frac{z}{6}\frac{[(1+I/J)-\frac{2}{z}\frac{D}{J}]^2}{1+\frac{4}{•3}\frac{I}{J}•},
\end{equation}
 with $0\leq \delta \leq \frac{z}{6}\frac{(1+I/J)^2}{1+\frac{4}{•3}\frac{I}{J}•}$. Notice that $\delta =0$ takes place, as aspected,  at the border line value $D/J=\frac{z}{2}(1+I/J)$ of the SIA parameter which separates the non-reentant and reentrant behaviors. Remarkably, the estimate \eqref{delta} for $\delta$ reproduces almost exactly the numerical findings of a sc lattice reported in Fig. 5 (panel (b)).
 
We now determine the asymptotic equation of the critical line in the regime $\Phi_c\gg 1$. By using  the known expansion of the Bose function into Eq. \eqref{phic} we get the solution
 \be\label{PHICASYMP}
 \Phi_c\simeq F_d(-1)\frac{T}{J(0)}\frac{1}{m}+\frac {1}{12}F_d(+1)\frac{J(0)}{T}m-\frac 12.
 \ee
Here $F_d(\pm1)$ are a particular case of the "lattice sums" $F_d(n)=\frac 1N\sum_{\vec k}(1-\gamma_{\vec k})^n$ where $\gamma_{\vec k}=J(\vec{k})/J(0)$ is the "structure factor".
Inserting \eqref{PHICASYMP} in  Eq.\eqref{BDIPHI}  we get the equation
 \be 
\label{mdiT}
\frac 23\left(\frac{J(0)}{T}\right)\frac{1}{F_d(-1)}\left\{1-\frac{1}{12}\left[1+F_d(-1)F_d(+1)\right]\right\}\left(\frac{J(0)}{T}\frac{m}{F_d(-1)}\right)^2=1,
 \ee
which relates $m$ and $T$ along the critical line as $H \rightarrow 0$.

On the other hand, from  Eqs. \eqref{BDIPHI} and \eqref{ASYMPCRITICAL} we find the relation $m\simeq \frac{H}{H_0}+O(\frac{1}{\Phi^3})$ where $H_0=J(0)+\frac{17}{18}I(0)-\frac{2}{3}D$. Then, to order of interest for $H\rightarrow 0$ we have the required critical line equation
\be
\label{TCDIH}
T_c(H)\simeq T_{0c}\left\{1-\frac{3}{16}\left[1+F_d(-1)F_d(+1)\right]\left(\frac{H}{H_0}\right)^2\right\},
\ee
where
\be
\label{TOC}
T_{0c}\equiv T_c (H=0) = \frac{2}{3}J(0)F_d^{-1}(-1).
\ee
Notice that, in our approximations, the parameters $I(0)$ and $D$ enter the problem only throught the parameter $H_0$.

Solving with respect to $H$ for the critical line as $T\rightarrow T_{0c}^-$ one find the representation
\be
\label{HCDIT}
\frac{H_c(T)}{H_0}\simeq \frac{4}{\sqrt{3}}\left[1+ F_d(-1)F_d(+1)\right]^{-1/2}\left(\frac{T_{0c}-T}{T_{0c}}\right)^{1/2}.
\ee

From Eqs.\eqref{TCDIH}-\eqref{HCDIT}, it is evident that all the critical lines in the ($H,T$)-plane terminate in the same critical point ($H_c=0, T_{0c})$. Finally, Eq. \eqref{HCDIT} allows us to obtain the  behavior of magnetization along the critical line as $m_c(T)\simeq  \frac{H_c(T)}{H_0}$ suggesting that $m_c(T)\rightarrow 0$ for $T\rightarrow T_{0c}^-$.

It is worth emphasizing that the presence of BQE coupling does not modify significantly the physics close to the QCP already known for the standard TXY model with SIA.

\section{Numerical findings and phase diagrams }
\label{secIV}

\begin{figure*}[bt]
\label{FIG.1}
\includegraphics[width=0.32\textwidth]{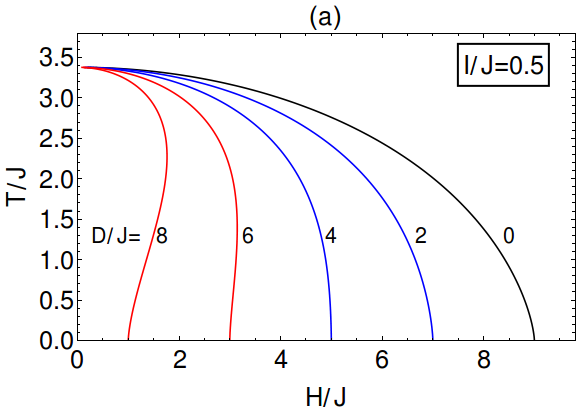} 
\includegraphics[width=0.32\textwidth]{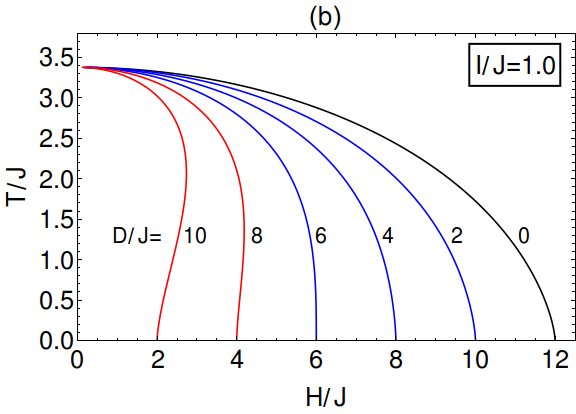} 
\includegraphics[width=0.32\textwidth]{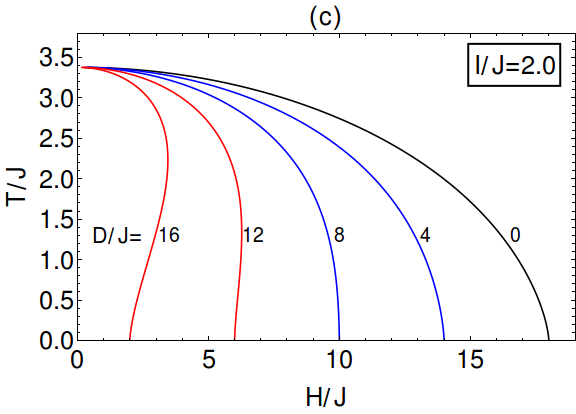}
\includegraphics[width=0.32\textwidth]{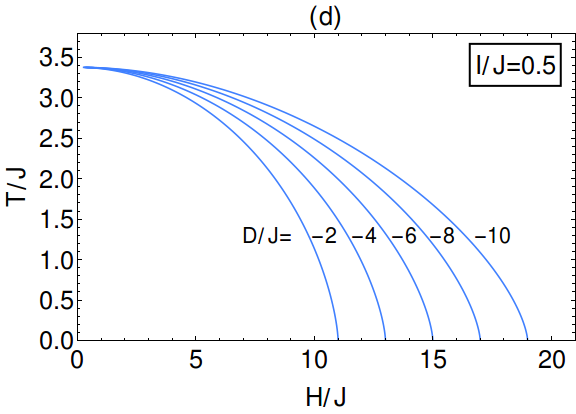}
\includegraphics[width=0.32\textwidth]{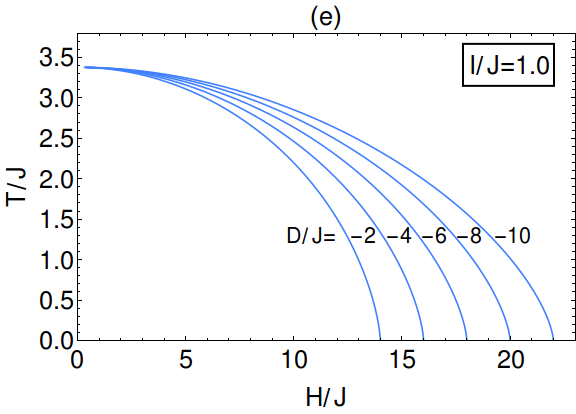}
\includegraphics[width=0.32\textwidth]{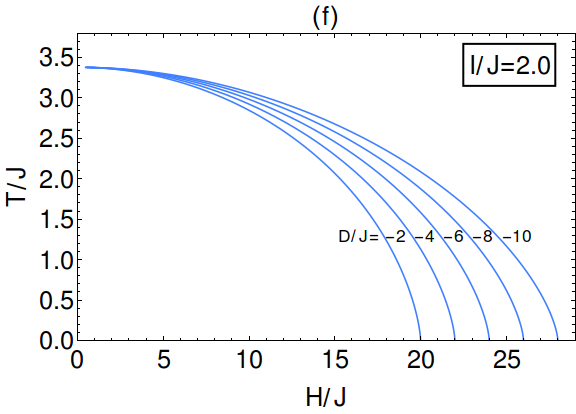}

\protect\caption{ { Critical lines in the $(H/J,T/J)$-plane with fixed $I/J$ by variation of $D/J$ for easy-axis (panels (a)-(c)) and easy-plane (panels (d)-(f)) SIA. Their behavior in the most interesting easy-axis regime indicates that the increasing $I/J$  reduces or destroys the reentrant structure of the critical lines. In the easy-plane case (panels (d)-(f)), for corresponding competing parameters no reentrances are present. } }
\end{figure*}
\begin{figure*}[bt]
\label{FIG.2}
\includegraphics[width=0.46\textwidth]{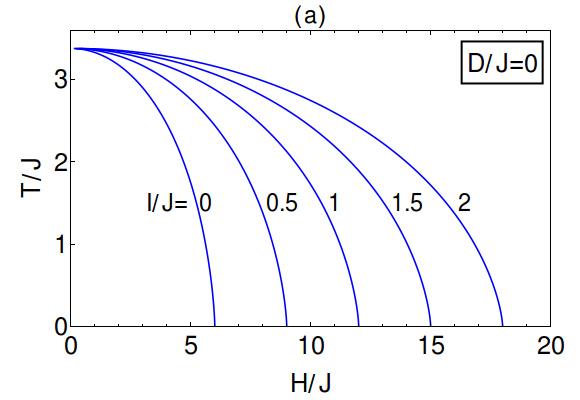} 
\includegraphics[width=0.46\textwidth]{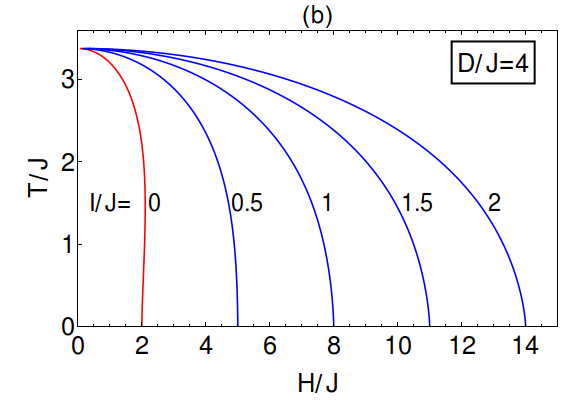} \\
\includegraphics[width=0.46\textwidth]{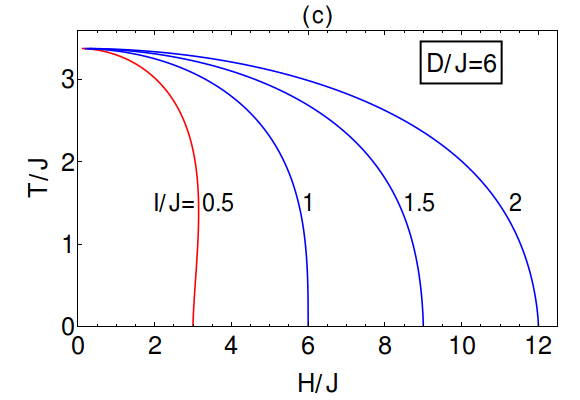} 
\includegraphics[width=0.46\textwidth]{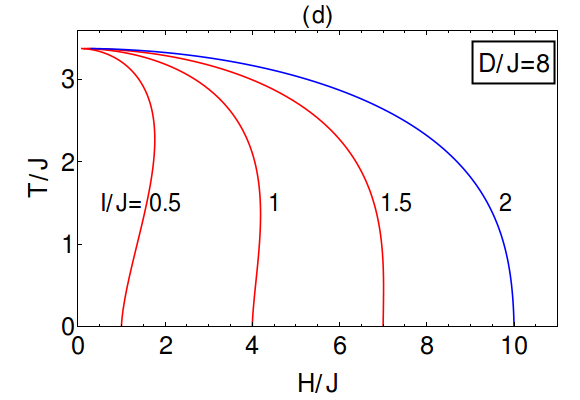} 
\protect\caption{Critical lines in the $(H/J,T/J)$-plane for fixed $D/J (=0, 4, 6, 8)$ for several values of the BQE adimensional coupling $I/J$.  In absence of SIA (panel (a)) the critical lines show a conventional behavior. In the panel (b) a reentrant behavior occurs in absence of BQE (red curve). The plots show that increasing $I/J$ reduces or destroys the  reentrant behavior  and a stronger easy axis SIA is required to observe it when the BQE becomes dominant (see panels (c) and (d)). }
\end{figure*}
\begin{figure*}[bt]
\label{FIG.3}
\includegraphics[width=0.46\textwidth]{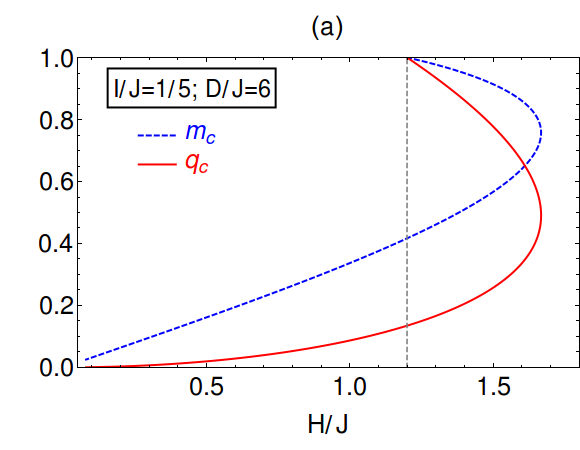} 
\includegraphics[width=0.46\textwidth]{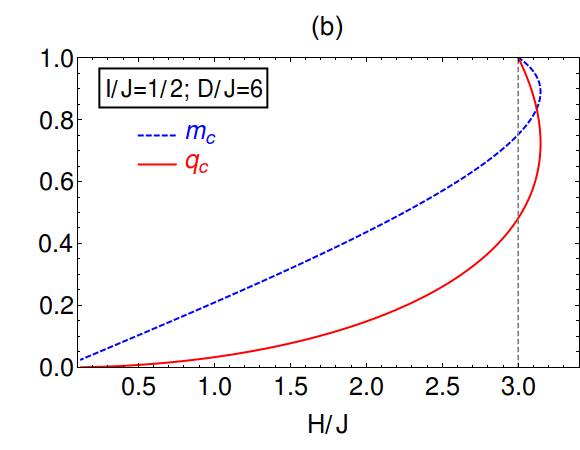}  \\
\includegraphics[width=0.46\textwidth]{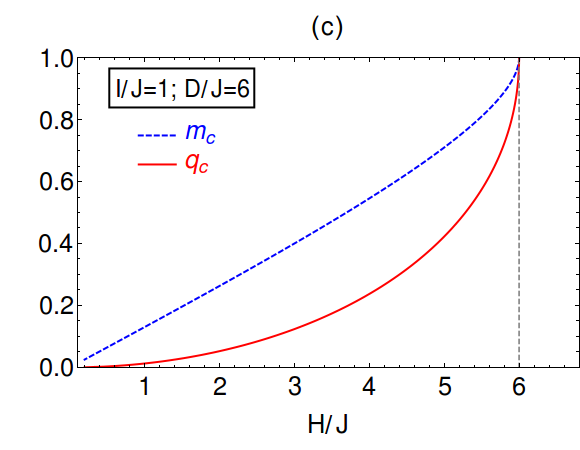} 
\includegraphics[width=0.46\textwidth]{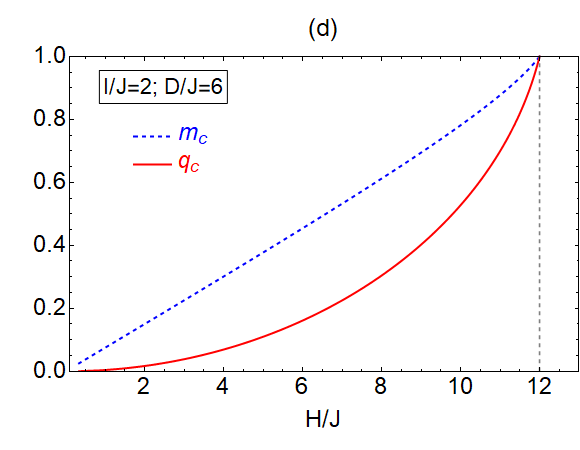} 
\protect\caption{Plots of magnetization, $m_c$, and quadrupolar order parameter, $q_c$, along critical lines as functions of $H/J$,  fixing the SIA parameter to $D/J=6$ and choosing different values for the BQE coupling $I/J$. In all the figures the value of $H_c$ at the QCP is marked by a dashed line. The double values of $m_c$ and $q_c$, observed when $H>H_c$,  signal the existence of a reentrance (panels (a) and (b)). In the panel  (c) the choice $I/J=1$, for $D/J=6$, selects the critical line with vertical tangent in $H_c$. A regular behavior is obtained for higher values of $I/J$ (see panel (d) where $I/J=2$).}
\end{figure*}
\begin{figure*}[bt]
\label{FIG.4}
\includegraphics[width=0.45\textwidth]{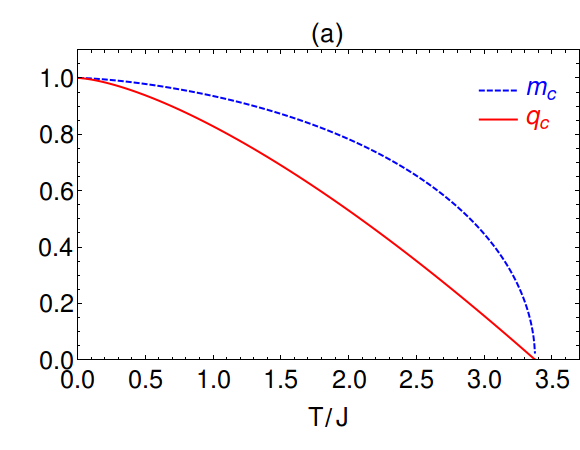} 
\includegraphics[width=0.45\textwidth]{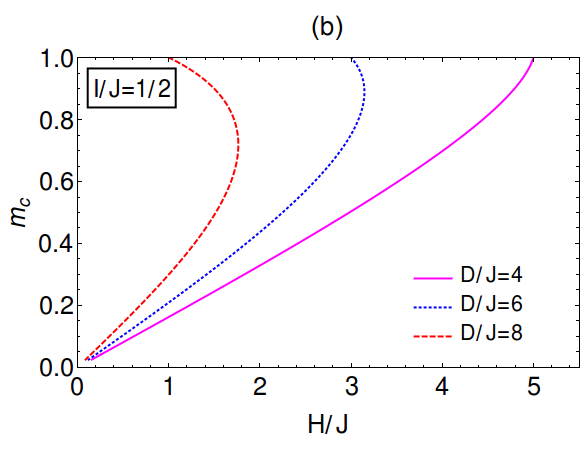} \\
\includegraphics[width=0.45\textwidth]{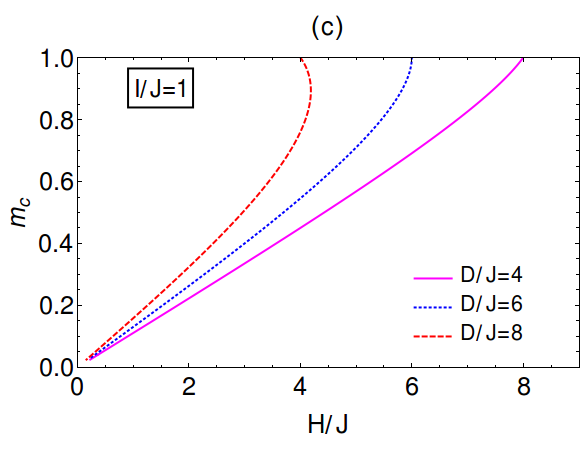} 
\includegraphics[width=0.45\textwidth]{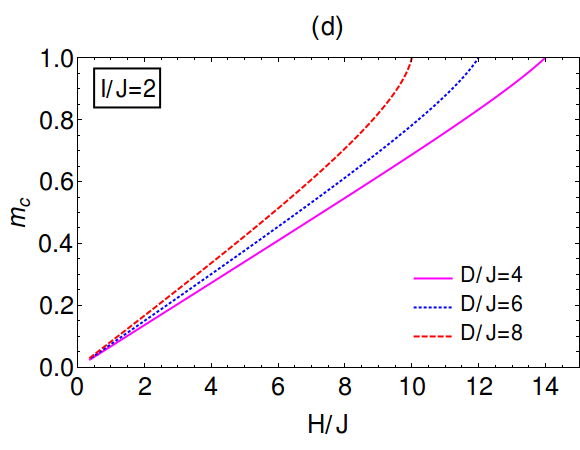} 
\protect\caption{ Panel (a) shows the behavior of $m_c$ and $q_c$ as functions of $T/J$; note that they do not depend on $I/J$ and $D/J$. In contrast, the behavior of $m_c$ as a function of $H/J$ (panels (b)-(d)) does depend on $I/J$ and $D/J$ and signals that the reentrance becomes, for fixed $I/J$, more and more marked increasing $D/J$. Similar resuls are valid also for $q_c$.}
\end{figure*}
\begin{figure*}[bt]
\label{FIG.5}
\includegraphics[width=0.46\textwidth]{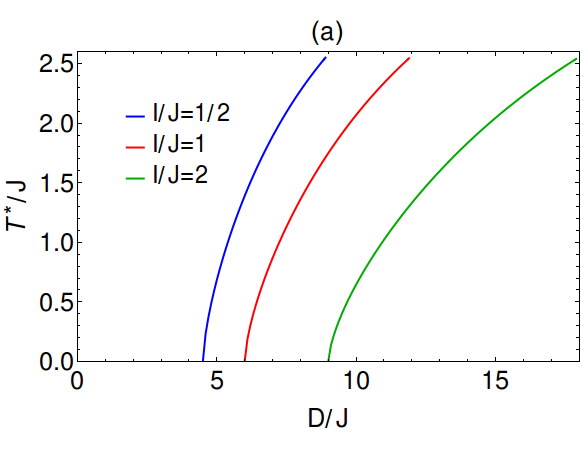} 
\includegraphics[width=0.46\textwidth]{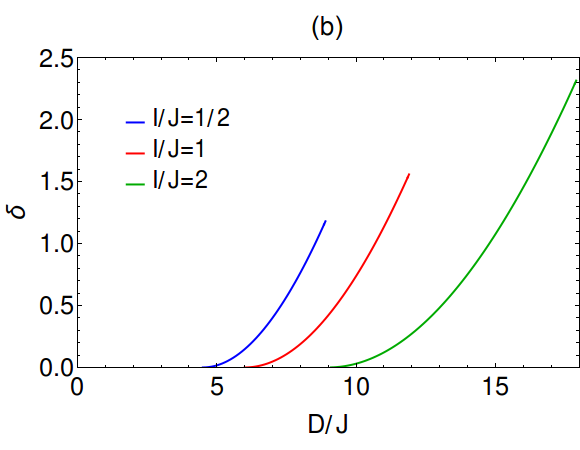}  \\
\includegraphics[width=0.46\textwidth]{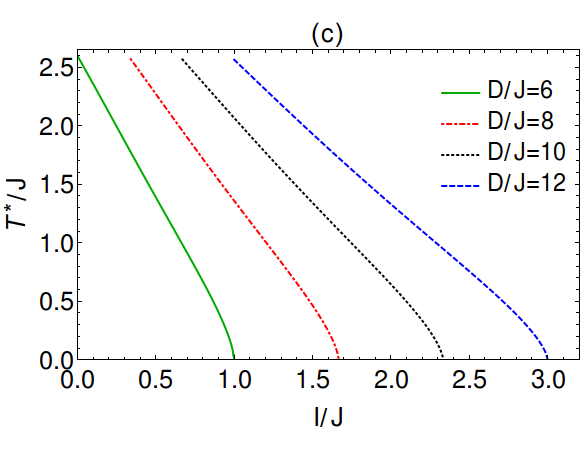} 
\includegraphics[width=0.46\textwidth]{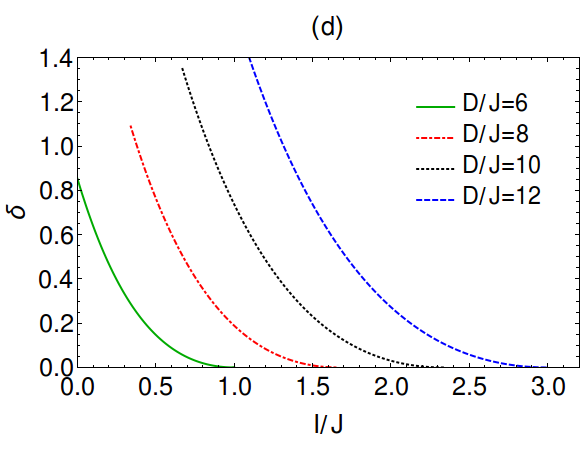} 
\protect\caption{ Plots of $T^*/J$ and  of the maximum amplitude of reentrances $\delta $ as function of $D/J$ for three values of $I/J$  (panels (a) and  (b)); in panels (c) and (d ) the same quantities are plotted as functions of $I/J$ for several values of $D/J$. The curves are obtained when reentrant behavior occurs, hence for $3(1+I/J) < D/J <6 (1+I/J)$ (panels (a) and (b)) or, equivalently, for  $D/6J -1 < I/J < D/3J-1$ (panels (c) and (d)). The lines in the panels (a) and (c) correspond to the critical points with coordinates $(H^*, T^*$ at which the critical lines have a vertical tangent. }  
\end{figure*}

The analytical estimates obtained in the Sec. III within the ACD   suggest an interesting but qualitative scenario about the effect of the BQE coupling on the phase diagram of the anisotropic quantum spin model \eqref{HAM}. Since the problem to solve analytically and exactly the self-consistent equations \eqref{phi}-\eqref{emme} for magnetization $m$ is prohibitive, we have solved them numerically obtaining  the required critical lines in the $(H,T)$-plane varying the competing parameters $D/J$, and $I/J$. We have focused, for the numerical  data, on a simple cubic lattice ($d=3, z=6$) but similar results can be derived for other cubic lattices.

Some phase diagrams for cases of easy-axis $(D>0)$ and easy-plane $(D<0)$ single-ion anisotropies (SIAs) are presented in Fig.1  for fixed ratio $I/J$ by variation of the reduced SIA parameter $D/J$. In the panels (a), (b) and (c), we have plotted three typical groups of critical lines for increasing values of the BQE and different values of the reduced SIA parameter $D/J$. From these plots it is evident that, in a certain range of easy-axis SIA parameter (estimated to be $3(1+I/J) < D/J < 6(1+I/J)$), reentrant critical lines occur. Beyond this range including the easy-plane case (panels (d)-(f)), and  consistently with previous analytical estimates, only regular  critical lines take place. 
Moving from the planel $(a)$ to panel $(c)$ it is also evident that,  increasing   the biquadratic to bilinear exchange ratio $I/J$, a crossover occurs to a regime where the reentrant behavior in the phase diagram is reduced or destroyed  and one needs to consider higher values of the reduced easy-axis SIA parameter $D/J$ for observing reentrancies. 

In the next figures  we will focus on the most interesting easy-axis regime.
In Fig.2, panels $(a)-(d)$, we present similar plots but with fixed $D/J (= 0, 4,6, 8)$ and by variation of $I/J$. Consistently with the results in Fig.1, the reentrance of the critical lines  reduces or disappears when biquadratic to bilinear ratio $I/J$ increases.  
Observe   that  in panels (c) and (d) it is lacking the critical line for $I/J=0$. Infact for $D/J\geq 6$ we are in the regime in which there is no critical line in absence of BQE.

It is worth emphasizing that the effect of BQE on the conventional magnetic-field-induced quantum criticality in  planar ferromagnets, when the crystal-field anisotrpy is absent, has not been explicitly explored up to now in literature.

Fig.3, panels (a)-(d), shows plots of magnetization $m$ and of the so-called quadrupolar order parameter $q = 3\langle (S^z)^2 \rangle - 2$ as functions of $H/J$   valued along the corresponding  critical lines. In this set of curves we have  fixed  $D/J = 6$ and  studied the effect of the variation of $I/J$. In the panels (a) and (b), the double values of $m_c$ and $q_c$ for fixed $H$ above the critical  field $H_c$ provide an alternative clear signal for the existence of reentrances in the phase diagram. The crossover to a regular critical line is observable for the particular couple of values $D/J = 6, I/J=1$ in the panel (c) but  growing $I/J$, a non-reentrant critical behavior is displayed (see panel (d)).  This feature may be an useful tool for selecting reentrant critical lines in possible experimental investigations. 

Additional information about $m_c$ and $q_c$ are shown in Fig. 4. In the panel (a) 
they are  plotted as functions of $T/J$ and the  panels (b)-(d) present the behaviors of $m_c$  as function of $H/J$ varying $D/J$ and $I/J$. Similar findings are true for $q_c$ but we omit them for for avoiding unuseful confusion. We see that in the case (a) no dependence  on the parameters $D/J$ and  $I/J$ occurs, consistently with the analytical estimate \eqref{emmeci}. In contrast the behavior as functions of $H/J$ appears sensibly dependent on both  $D/J$ and  $I/J$ with a clear evidence of reentrances and of their reduction increasing  $I/J$.

Finally, in  Fig. 5 we plot  the quantities  $T^*/J$ and $\delta = (H^* - H_c)/J$ as functions of $D/J$ (panels (a) and (b)) and  of $I/J$ (panels (c) and (d)) in the reentrance region. Here $H^*$  denotes the vertical  tangent to the critical lines in the $(H,T)$-plane and $T^*$ is the  corresponding temperature. In particular the  figures in panels (a) and (b) provide a clear  quantitative representation for the reduction or distruction of reentrances increasing the BQE coupling. 

It is worth emphasizing that the estimates obtained analytically in Sec. III appear qualitatively, and sometimes quantitatively, consistent with the present numerical findings.

\newpage

\section{CONLUDING REMARKS }

In this paper we have investigated the effect of the BQE interaction on the phase diagram of a $d$-dimensional spin-1 ferromagnetic XY model with a SIA in the presence of a transverse magnetic field. We have employed the two-time GF method by using the ACD  treatment for the crystal-field anisotropy-like terms in the EM for the GF of interest. The exchange higher order GFs have been decoupled at the RPA level. A unified and rich picture has been  derived for short-range interactions with dimensionalities $d>2$ by using analytical estimates and numerical calculations for easy-plane and easy-axis SIA. The numerical data, collected in Figs.1 - 5, refer to a $sc$ lattice ($z=3$) but similar findings easily follow for $bcc$ ($z=8$) and $fcc$ ($z=12$) lattices.

The main results arising from the previous  scheme indicate that, in presence of SIA: (i) for small BQE parameter the scenario appears quite similar to that already known  when only the conventional bilinear exchange  interaction is active \cite{noi14b}, \cite{noi17},\cite{mtm17}; (ii) increasing the BQE parameter  tends to destroy the reentrant structure of the critical lines which occurs when the bilinear coupling is dominant. 

From our findings, the physical origin of the reentrances in the phase diagram of the spin model \eqref{HAM} remains an open question. Some insights could be extracted from the illuminanting general arguments presented in Refs. \citep{diep3}, \citep{diep4} for exactly solved magnetic models with frustration. Nevertheless, since for our model no exact results are available, it is not easy to apply such arguments. Anyway the following qualitative considerations about the onset of reentrances can be made.
The phase diagrams in the $(H,T)$-plane for fixed $D/J$ and  increasing the exchage ratio $I/J$, or viceversa, are shown in Figs. 1 and 2. Reentrant phase transitions appear  for appropriate values of the easy-axis SIA parameter and the exchange ratio $I/J$. When the temperature is lowered along a vertical line, one observes first a disorder to order transitions  and  then, further lowering the temperature, an order to disorder transition. Besides, increasing $H$ around the quantum critical magnetic field $H_c(D)$, one crossovers from a situation where a single phase transition occurs to one where two phase transitions are present for each fixed $H$ up to a limit value $H^*$ where the vertical line is tangent to the critical line. Both these features are characteristic of reentrant phenomena. Quantum effects and crystal-field anisotropy compete in the ordering process, but the two effects do not affect in the same way the phase diagram. 
Indeed, lowering the temperature from the disordered phase, quantum fluctuations dominate and a conventional quantum spin transition occurs. Further lowering  the temperature, the SIA contributes mainly and the reentrance to the disordered phase emerges. In any case the easy-axis nature of the SIA is the responsible of the  reentrances for  values of $D$ in the interval  ($J(0)+I(0))/2 < D < (J(0) + I(0)$). Increasing $D$ reduces the  quantum critical magnetic field $H_c(D)$, the critical lines blend and   the reentrances become more and more pronounced together with the  reduction of the in-plane ordered phase. Beyond this range, i.e. for $0 \leq D \leq ( J(0)+I(0))/2)$ and in the case of easy-plane SIA $(D<0)$ where $H_c(D)$ enlarges when $|D|/J$ increases, no reentrances are present in the phase diagrams. Of course, previous arguments are not exhaustive and further investigations are needed to clarify the mechanism of the reentrant transitions close to the QCPs of the transverse anisotropic XY model with BQE and SIA.”  

In view of recent literature \cite{noi14b}, \cite{noi17}, \cite{mtm17} and references therein, we speculate that our ACD analysis provides a reasonable physical scenario which may give useful insights on the structure of the  phase diagram near magnetic-field-induced quantum criticality. Quantitative improvements could be obtained by employing Callen-like decouplings \cite{CallenM} in  the exchange terms \cite{Kumar, CHADDA1,SIMING}, allowing estimates of the correlation effect between the transverse components of spins on different sites. However, this procedure complicates  the mathematical calculations  without a substantial change of the qualitative physics arising from the simplest and controllable RPA.
One can try to avoid  the ACD by treating exactly the SIA-like terms in the EMs according to the original idea by Devlin \cite{devlin} but the basic equations become too complicate also for numerical calculations. However, in absence of BQE couplings, it has been recently shown \cite{noi14a} that the simplest ACD approach turns out to be rather efficient in investigating the criticality close  to the magnetic-field-induced QCP in planar ferromagnets. In particular, it allowed to distinguish and study in detail  two types of criticality close to a QCP: conventional quantum criticality, when a regular  critical line takes place and a non conventional quantum criticality, when a reentrant critical line is present. On this ground, we believe that the same situation occurs also for the model \eqref{HAM} which includes the BQE. So one can describe its conventional and non-conventional quantum criticalities by means of the method developed in Ref. \cite{noi14a}.

Remarkably, in the present paper we have shown that, by variation of $D$, $I$,  and $J$  a well defined range of SIA parameter exists where the reentrant critical lines occur and that increasing the BQE parameter tends to reduce or destroy their reentrant structure. This suggests that a BQE-induced crossover from a non conventional quantum criticality to a conventional one occurs  increasing the BQE.

At our knowledge there is no clear experimental evidence of spin-1 three-dimensional compounds with competing effects of thermal and quantum fluctuations, single-ion anisotropy and biquadratic interactions,  for which the Hamiltonian \eqref{HAM} may be applied for searching reentrances in their phase diagrams. A class of advanced magnetic materials, which seem to present these competing ingredients,  and for which the quantum anisotropic XY model \eqref{HAM} could be applied, are  the three-component systems with weak interplane interactions as $K_2CuF_4$, $(CH3NH3)_2CuCl_4$, $BaCo_2(AsO4)_2$  \cite{ivanov}, \cite{YU}. Besides, with the inclusion of a single-site orthorombic anisotropy, the quantum XY model \eqref{HAM} is believed to be appropriate for some magnetic materials of the $MnCl_2 \cdot 4H2O$ type \cite{ciep,Ma}. Interestingly, there are also systems with spin $S\geq 1$ magnetic ions whose  properties can be properly described, at least in principle, by Hamiltonians with XY symmetry which involve comparable bilinear and biquadratic exchange interactions. In such cases a wide variety of unusual magnetic properties have to be still explored. Thus, it is of increasing theoretical and experimental interest to investigate the low-temperature properties of the spin-1 XY model in a transverse field, including a biquadratic exchange interaction and a single-ion anisotropy, which exhibits also a field-induced QPT. 

In conclusion, since to our knowledge no explicit studies have been achieved in the past for exploring the effect of BQE interactions around a magnetic-field-induced QCPs in presence or in absence of SIA, we believe that the present study may stimulate further theoretical and experimental investigations about complex magnetic materials with crystal-field anisotropy which exhibit quantum phase transitions and reentrant phase diagrams close to the QCP.


\newpage







\begin{thebibliography}{00}


\bibitem{Chen} H. H. Chen, P. Levy, Phys. Rev. B 7 (1973) 4284.
\bibitem{Micn} R. Micnas, J. Phys. C: Solid State Phys. 9 (1976) 3307.
\bibitem{Chad} G.S. Chaddha, A. Sharma, J. Magn. Magn. Mater. 191 (1999) 373 and references therein.
\bibitem{tya} S.~V. Tyablikov, {\it Methods in the Quantum Theory of Magnetism} (Plenum Press, New York, 1967).
\bibitem{nolting} W. Nolting and A. Ramakanth ``Quantum theory of Magnetism", Springer-Verlag, Berlin, Heidelberg (2009).
\bibitem {zubarev} D.N. Zubarev, Usp. Fiz. Nauk 71, 71 (1960) (Sov. Phys. Usp, 3, 320 (1960).
\bibitem {diep1} H.T. Diep, Phys. Rev. B 91, 014436 (2015), and references therein.
\bibitem {diep2} Sahbi El Hog and H.T. Diep, J. Magn. Magn. Mater., 400, 276 (2016).
\bibitem {diep3} H.T. Diep , Journal of Science: Advanced Materials and Devices 1 (2016) 31, and references therein.
\bibitem{Ega} T. Egami, B.~V. Fine, D. Parshall. A. Subedi, and D.J. Singh, Adv. Cond. Matt. Phys. 2010, ID 164916. 
\bibitem{Hirs} P.J. Hirschfeld, M.M. Korsjunov and I.I. Mazin, Rep. Progr. Phys. 74 (2011) 124508. 
\bibitem{Wis} A. L. Wysocki, K.D. Belashchenko, and V.P. Antropov, Nature Phys. 7 (2011) 485 and references therein. 
\bibitem{Wis2} A.L. Wysocki, K.D. Belashchenko, L. Ke, M. van Schilfgaarde and V.P. Antropov, J. Physics: Conference Series 449 (2013) 012024.
\bibitem{And} E.C. Andrade, M. Brando, C. Geibel, and M. Vojta, Phys. Rev. B 90 075318 (2014).
\bibitem{Bra} M. Brando, D. Belitz, F.M. Grosche, and T.R. Kirkpatrik, Rev Mod. Phys 88 (2016) 025006 and references therein.
\bibitem{sachdev} S. Sachdev, {\em Quantum Phase Transitions} (Cambridge University Press, Cambridge 2011)
\bibitem{noi07}  M. T. Mercaldo, L. De Cesare, I. Rabuffo, A. Caramico
D'Auria, Phys. Rev. B, 014105 (2007) 75 .
\bibitem{noi10} L. S. Campana, L. De Cesare, U. Esposito,  M. T. Mercaldo,  I. Rabuffo, Phys. Rev. B, 024409 (2010) 82 .
\bibitem{Chad2} G.S. Chadda and S.M. Zheng, J.Magn. Magn. Mater. 152 (1996) 152.
\bibitem{Chen2} K.G. Chen, H.H. Chen, C.S. Hsue, F.Y. Wu, Physica 87A (1977) 629.
\bibitem{Chen3} K.G. Chen, H.H. Chen, C.S. Hsue, F.Y. Wu, Physica 87A (1978) 526.
\bibitem {Zuk} M. Zukovic, T. Idogaki, K. Takeda, Physica B 304 (2001) 18.
\bibitem {Zuk2} M. Zukovic, T. Idogaki, Physica B 324 (2002) 360. 
\bibitem{Dut} A. Dutta, G. Aeppli, B.K. Chakrabarti, U. Divakaran, T. F. Rosenbaum, D. Sen,    "Quantum Phase Transitions in Transverse Field Spin Models: From Statistical Physics to Quantum Information", Cambridge University press 2015.
\bibitem{lima} L. S. Lima, A.S. T. Pires,   Eur. Phys. J.  B 70 (2009) 335.
\bibitem{noi14a} M.T. Mercaldo, I. Rabuffo, L. De Cesare, A. Caramico D’Auria, J. Magn. Magn. Mater., 364 (2014) 85.
\bibitem{noi14b} M.T. Mercaldo, I. Rabuffo, L. De Cesare, A. Caramico D’Auria, J. Phys.: Conf. Ser. 529 (2014) 012019.
\bibitem{noi15}  I. Rabuffo, A. Caramico D'Auria, L. De Cesare, M.~T. Mercaldo,  J. Magn. Magn. Mater. 382 (2015) 237.
\bibitem{noi16} M.T. Mercaldo, I. Rabuffo, L. De Cesare, A. Caramico D’Auria, J. Magn. Magn. Mater., 403 (2016) 68.
\bibitem{ACD} F.B. Anderson and H. Callen, Phys. Rev.  136 (1964) A1068.
\bibitem{froe} P. Fr\"obich and P.~J. Kunz, Phys. Rep. 432 (2006) 223 and references therein.
\bibitem{tanaka} M. Tanaka, Y. Kondo, Progr. Theor. Phys.   48 (1972) 1815.
\bibitem{noi06} I. Rabuffo, M.T. Mercaldo, A. Caramico D'Auria, L. De Cesare, Phys. Lett. A  356 (2006) 174.
\bibitem{noi08} M.~T. Mercaldo, A. Caramico D'Auria, L. De Cesare, I. Rabuffo, Phys. Rev. B 77 (2008) 184424.
\bibitem{CallenM} H. B. Callen, Phys. Rev. 130 (1963) 890.
\bibitem{noi13} M.~T. Mercaldo, I. Rabuffo, L. De Cesare, A.Caramico~D'Auria, Eur. Phys. J. B 86 (2013) 340.
\bibitem{diep4} H.T. Diep and H. Giacomini, Chapter I, in “Frustrated Spin Systems”, Ed. H.T. Diep, World Scientific (2013).
\bibitem{MW}   N. D. Mermin, H.  Wagner   Phys. Rev. Lett.  17  (1966) 1133.
\bibitem{Kelley} C. T. Kelley, "Iterative methods and non-linear equations", Society for Industrial and applied Mathematics, Phyladelfia, 1995.
\bibitem{noi17} M.~T. Mercaldo, I. Rabuffo, L. De Cesare, A.Caramico~D'Auria,  J. Magn. Magn. Mater. 439 (2017) 333.
\bibitem{mtm17} I. Rabuffo, L. De Cesare,  A. Caramico D'Auria, M.T. Mercaldo,  https://doi.org/10.1016/j.physb.2017.10.064
\bibitem{Kumar} V. Kumar and K.C. Sharma, Prog. Theor. Phys. 56 (1976) 801.
\bibitem{CHADDA1} G.S. Chaddha, G.S. Kalsi, Phys. Stat, Sol. (b) 151 (1989) 283.
\bibitem{SIMING} Z. Siming, J. Phys. Chem. Solids 47 (1986) 255.
\bibitem{devlin} F. Devlin, Phys. Rev. B (1971) 136.
\bibitem {ivanov} B.A. Ivanov and A.K. Kolezhuk, Low Temp. Phys. 21, 760 (1995).
\bibitem{YU} Yu. A. Fridman, O.A. Kosmachev, and F.N. Klevets, Low Temp. Phys, 32 (2006) 1.
\bibitem{ciep} M. Cieplak, Phys. Rev. B 15, 5310 (1977).
\bibitem{salinas} W. Figueiredo And S. R. Salinas, Physica B 124, 259 (1984).
\bibitem{Ma} Ma  Yu-qiang and W. Figueiredo, Phys. Rev. 55, 5604 (1997).

\end{thebibliography}
\end{document}